%% file: PBH-21cm.tex
\documentclass[aps,prd,10pt,twocolumn,nofootinbib,floatfix]{revtex4}
\usepackage[title]{appendix}
\usepackage{epsfig}
\usepackage{bm}
\usepackage{latexsym}
\usepackage{natbib}
\usepackage{url}
\usepackage{dcolumn}
\usepackage{color}
\usepackage{amsfonts,amssymb,amsmath}
\usepackage{graphicx,epsfig}
\usepackage{psfrag}
\usepackage{subfigure}
\usepackage{hyperref}
\hypersetup{colorlinks=true}
\usepackage{mathtools}
\usepackage{enumitem}
\usepackage{float}
\usepackage{mathrsfs}
\usepackage{xcolor}
\hypersetup{colorlinks,  citecolor=red,  linkcolor=red,  urlcolor=blue}
\usepackage{comment}
\usepackage{tikz-feynman}
\usepackage{graphicx}

\usepackage{mathtools}
\numberwithin{equation}{section}

\usepackage{mathrsfs}
\usepackage{gensymb}
\begin{document}

\title{Impact of Primordial Black Holes Induced Neutrinos on the Cosmic 21-cm Brightness Temperature}

\author{
Prabhav Singh$^a$\footnote{24RPH003@spt.pdpu.ac.in},
Mansi Dhuria$^a$\footnote{Mansi.dhuria@sot.pdpu.ac.in},
Gaurav Goswami$^{b,c}$\footnote{gaurav.goswami@ahduni.edu.in}
}

\affiliation{
$^a$Department of Physics, School of Energy Technology, Pandit Deendayal Energy University (PDEU), Gandhinagar-382426, Gujarat, India\\
$^b$Division of Mathematical and Physical Sciences, School of Arts and Sciences, Ahmedabad University, Ahmedabad-380009, Gujarat, India\\
$^c$International Centre for Space and Cosmology, Ahmedabad University, Commerce Six Roads, Navrangpura, Ahmedabad - 380009, India
}

\begin{abstract}
\noindent 

We study the impact of neutrinos emitted from evaporating Primordial Black Holes (PBHs) on the global 21-cm absorption signal during the dark ages and pre-reionization epochs. PBHs emit neutrinos over a wide energy range through Hawking evaporation. We investigate the possibility that radiative scattering between these neutrinos and the Cosmic Neutrino Background (C$\nu$B) generates secondary photons, leading to additional heating of the neutral hydrogen gas. This modifies the thermal history of the intergalactic medium and increases the global 21-cm brightness temperature relative to the standard cosmological prediction. Using the absorption feature at redshift $z\simeq17$, we derive new constraints on the PBH fraction for PBH masses in the range $10^{15}\mathrm{g}\lesssim m_{\rm BH,0}\lesssim10^{25}\mathrm{g}$. We further use existing PBH limits to constrain neutrino self-interaction couplings over a broad range of mediator masses. Our analysis complements previous studies that focused on direct photon injection from PBH evaporation and highlights the importance of neutrino-induced effects within a multimessenger framework for probing PBHs and beyond-standard-model neutrino interactions.


\end{abstract}

\maketitle
\section{Introduction}
Over the years, our knowledge of the universe in its infancy as well as current times has increased tremendously by probing the Cosmic Microwave Background (CMB) and the observations from current local structures. However, we still lack sufficient understanding of the intermediate universe when the last scattered photons were emitted from the CMB until the time of formation of the first stars. The reason for the lack of understanding during that epoch is due to the universe being largely dominated by the neutral hydrogen (HI), hence having no absorption or emission of photons which can be detected. One of the only ways by which we can probe the intermediate universe from redshift $z\sim 1000$ to $z\sim 6$ is by observing the 21 cm photons emitted from the transition of the two hydrogen states. The time immediately following the CMB emission is known as the period of ``dark ages" in which the universe was dormant with no source producing photons except for the transition of hydrogen states. The era of ``cosmic dawn" is the period after the dark ages in which the universe became cool enough for the first stars and galaxy structures to form and is followed by the Epoch of Reionization (EoR) in which the universe transitions into fully ionized, resulting in the re-ionization of the universe. Hence for dark ages, cosmic dawn, as well as EoR, the study of emission and absorption of photons from neutral hydrogen is necessary to study the intermediate universe. Presently, the intermediate universe is mainly probed through indirect observations from signals coming from CMB Thomson scattering, high redshift quasars as well as Lyman-$\alpha$ emitters\cite{fan2002evolution,Ciardi:2003hg,Vishrut-paper}. Also, numerous ongoing and future radio telescopes like LEDA~\cite{LEDA}, REACH~\cite{reach_homepage}, Murchison Widefield Array (MWA)~\cite{MWA}, Square Kilometre Array (SKA)~\cite{SKA-white-paper,Koopmans:2015sua}, upgraded Giant Meter-wave Radio Telescope (uGMRT)~\cite{GMRT}, Hydrogen Epoch of Reionization Array (HERA)~\cite{HERA,HERA:2022wmy}, Low Frequency Array (LOFAR)~\cite{LOFAR} and EDGES~\cite{EDGES} are trying to detect the neutral hydrogen signal emitted during that period. 
In recent years, the detection of an anomalously deep global 21-cm absorption feature by the EDGES experiment in 2018 generated significant interest in exploring the impact of new physics on the 21-cm signal. Since then, the global 21-cm line has emerged as a sensitive probe of several cosmological and astroparticle physics scenarios, including dark matter decay and annihilation processes, constraints on cosmological parameters such as the Hubble constant, and signatures of exotic early-universe phenomena such as Primordial Black Holes (PBH) and non-standard inflationary models. Although the EDGES result was later challenged by the SARAS observations at 95\% significance level~~\cite{Singh:2021mxo}, interest in the 21-cm signal remains strong due to the prospect of improved measurements from current and future-generation radio experiments~~\cite{21-cm-SKA,DM-detector,SKA-white-paper,Dhuria:2023itq}.

Consequently, several non-standard mechanisms that can modify the global 21-cm brightness temperature have been extensively investigated. These include deviation in the brightness temperature due to dark matter annihilation, non-standard neutrino self-interactions~\cite{Barkana:2018qrx},  inhomogeneous X-ray heating from DM decay~\cite{Pritchard:2006sq}, presence of early radio background~\cite{Feng:2018rje}, among others. 
One such non-standard process is the production of secondary photons through the scattering of neutrinos with the Cosmic Neutrino Background (C$\nu$B), which can affect the global 21-cm brightness temperature. Neutrinos spanning a wide range of energies can originate from several astrophysical and cosmological sources, including dark matter decay, PBH evaporation, and supernova explosions. In this work, we investigate the impact of neutrinos emitted through PBH evaporation on the global 21-cm signal.

Primordial Black Holes are black holes believed to have formed very early in the universe from the collapse of the overdense region~\cite{OUR-PAPER,HEandUHEchina,NonCDMpaper,SevenhintsPBH,PBH_book,carr1974black,hawking1974blackexplosion}. As the region collapsed to form PBH, it also started to accrete nearby matter, hence also being the seeds for galaxy formation. 
PBHs can also accrete surrounding matter and have been proposed as possible seeds for structure and galaxy formation. Depending on their mass, PBHs evaporate through Hawking radiation, emitting a wide range of particles, including neutrinos.
There have been numerous studies involving PBH evaporation as a source of energy injection capable of altering the spin temperature, and hence the brightness temperature, of the universe. These include Hawking radiation from light PBHs and emission from accreting solar-mass PBHs, which influence the inhomogeneous energy injection and consequently modify the brightness temperature and the 21-cm power spectrum~\cite{PBH-21cm-Sun}. Other studies have investigated the impact of Active Galactic Nuclei (AGN) and star-forming (SF) galaxies seeded by PBHs on the 21-cm signal at high redshift~\cite{PBH-21cm-Chatterjee}, explored PBH constraints in the mass range \(10^{10}\,\mathrm{g} \lesssim m_{\rm PBH} \lesssim 10^{14}\,\mathrm{g}\) from Hawking radiation of photons and \(e^\pm\) affecting the brightness temperature~\cite{PBH-21cm-mack}, as well as the injection effects from memory-burdened PBHs~\cite{PBH-21cm-Sarmah} and energy injection from \(\gamma\) and \(e^\pm\) emission from spinning and non-spinning PBHs~\cite{PBH-21cm-Pravin,PBH-21cm-saha,PBH-21cm-clark1,PBH-21cm-cang}.
Nonetheless, existing studies primarily focus on the direct photon emission from PBH evaporation as the main source of heating of the intergalactic medium (IGM), thereby modifying the global 21-cm brightness temperature. In this work, we instead investigate the role of neutrinos emitted from PBH evaporation, which can generate secondary photons through loop-level scattering processes. These photons subsequently heat the IGM gas and alter the thermal history relevant to the 21-cm signal. Using this mechanism, we derive improved constraints on the PBH abundance over a wide range of PBH masses. Additionally, for a viable range of PBH masses and PBH fractions, we also derive constraints on the neutrino self-interaction coupling and the corresponding mediator mass.

The paper is organized as follows: In section~\ref{21-cm-Basics}, we review the fundamentals behind the 21-cm physics and standard processes affecting the brightness temperature. In section~\ref{gas-temp-standard}, we study the evolution of the kinetic heating of the IGM gas from the prior standard process. In section~\ref{gas-temp-standard}, we study the evolution of the kinetic temperature of the IGM gas in standard cosmology and in the presence of exotic energy injection processes, respectively. In particular, we examine the modifications to the gas temperature and free electron fraction arising from effects such as gas heating induced by PBH evaporation.
In section~\ref{PBH-neutrinos}, we derive the differential neutrino flux emitted from PBHs through Hawking evaporation. This flux is subsequently used in sec.~\ref{PBH-energy-deposition}, where we compute the energy deposition into the IGM arising from neutrino interactions. In section \ref{results}, we present the impact of PBH-induced gas heating on the global 21-cm signal and derive updated constraints on the PBH abundance and self-interacting neutrino coupling during the cosmic dawn era. In section~\ref{Summary}, we summarize our results and discuss future directions.

\section{21-cm Global Signal}
\label{21-cm-Basics}

After the Cosmic Microwave Background, the universe became dominated by hydrogen atoms, which were formed from the recombination of electrons and protons. As there was no other stimulus in the universe for the photons to be emitted, the only photon emission was through the transition between the singlet and triplet states of neutral hydrogen. This happens when the spin of the electron transitions from parallel to anti-parallel (or vice-versa) alignment compared with that of the proton. Naturally, the energy of the hydrogen atom in the triplet state is greater than the energy in the singlet state. Hence, when hydrogen in the triplet state comes down to the singlet state through some interactions, a single photon is emitted. The energy difference between hydrogen in the triplet and singlet state is $5.9\rm \, \mu eV$, which corresponds to the wavelength of $\lambda = 21\,\rm cm$, which lies in the radio-wave region of the electromagnetic spectrum. Although the transition between the states is exceedingly rare, the high abundance of neutral hydrogen during that time implies that the 21 cm photons were nevertheless emitted in the universe. If we are able to probe these redshifted photons from the background photons of the universe, we might be able to provide some explanation of different interactions happening during the dark ages and cosmic dawn. To observe the emission of 21 cm photons, we first need to calculate the Brightness Temperature ($T_b$) of the universe, which is defined as the observed deviation in the blackbody emission spectra at the 21 cm wavelength, taking into account the cosmological redshift of the photons that were emitted during that time. In other words, it depends on the change observed from the theoretical blackbody emission framework, i.e., $T_b \propto I_{\nu}^{\rm observed} - I_{\nu}^{\rm CMB}$. In the standard cosmology, there are three major types of processes by which transitions between the triplet and singlet states can happen, thereby emitting the 21 cm photons. Those processes are the excitation or de-excitation of hydrogen gas due to the cosmic microwave background, emission of 21 cm photons from the kinetic heating of the gas, primarily from the collisions between the electrons, protons, and hydrogen gas with each other, and lastly the 21 cm photon emission through the resonant scattering of Lyman-$\alpha$ photons, also known as the Wouthuysen Effect. The brightness temperature can be quantified as~\cite{Pritchard-Loeb, Dhuria:2024zwh, Cosmo-low-freq-data1, 21-cm-SKA,Dhuria:2021lqs}
\begin{equation}
  \mathrm T_b = 27\,x_{\rm HI}\,\left[\frac{0.15}{\Omega_m}\frac{1+z}{10}\right]^{1/2}\left(\frac{\Omega_b\,h}{0.023}\right)\left(1-\frac{T_\gamma}{T_s}\right)\,{\rm mK},
  \label{T21-final}
\end{equation}
where, $x_{\rm HI}=n_{\rm HI}/n_{\rm H} = 1-x_{\rm e}$ is the fraction of neutral hydrogen in the universe with $x_e$ being the fraction of ionised hydrogen, $\rm T_\gamma=T_{CMB}$, $h=0.74$ is the reduced Hubble constant, $\Omega_b \approx 0.044,\, \Omega_m\approx0.29$ are the relic abundance of baryonic and total matter, $\rm z$ is the redshift and $T_s$ is the spin temperature which depends on different kinds of interaction for absorption or emission of 21 cm photon. Naturally, from the eq.~\ref{T21-final}, we can see that the brightness temperature is dependent on the fraction of neutral hydrogen present in the universe. So, if there is no neutral hydrogen in the universe, the brightness temperature will be zero. 

The spin temperature is an artificial non-physical parameter defined to describe the relative population of the two states. In standard cosmology, the spin temperature $T_s$ is governed by three main processes: coupling to the CMB radiation, atomic collisions, and Ly$\alpha$ interactions (Wouthuysen-Field effect)~\cite{Pritchard-Loeb, Dhuria:2024zwh, Cosmo-low-freq-data1,21-cm-SKA}. The resulting spin temperature is given by
\begin{equation}
    \mathrm T_s^{-1} = \frac{T_\gamma^{-1}+x_c\,T_k^{-1}+x_\alpha T_c^{-1}}{1+x_c+x_\alpha},
    \label{spin-temp}
\end{equation}

where $T_\gamma$ is the CMB temperature, $T_k$ is the kinetic temperature of the gas due to atomic collisions, with $x_c$ being the collisional coupling, and $T_c$ is defined as the color temperature arising due to the Wouthuysen effect with $x_\alpha$ being the coupling coefficients. The $T_\alpha$ and $T_k$ are coupled due to the recoil from repeated scattering. Hence, the spin temperature follows the kinetic temperature when sum of coupling coefficients $x_\alpha+x_c \gtrsim 1$ and follows C$\nu$B when $x_\alpha+x_c \ll 1$.

\subsection{Kinetic heating of gas}
In this subsection, we review the collisional coupling coefficient and its dependence on various physical parameters. As discussed above, spin-flip transitions in neutral hydrogen can occur through collisions between hydrogen atoms or with free electrons and protons. The corresponding collisional coupling coefficient for a species $i$ is given by~\cite{Pritchard-Loeb}
\begin{equation}
    x_c^i = \frac{C_{10}\, T_\star}{A_{10}\, T_\gamma} = \frac{n_i\,k_{10}^i}{A_{10}}\,\frac{T_\star}{T_\gamma},
\end{equation}
where $C_{10}$ is the collisional excitation rate, $A_{10} = 2.85\times10^{-15}\, s^{-1}$ is the spontaneous decay rate of the spin flip transition, $k_{10}^i$ is the specific rate coefficient arising from the collision of hydrogen atom with species "$i$" resulting in spin de-excitation in the units ${\rm cm}^3\,{\rm s}^{-1}$ and $T_\star = T_{21} \equiv \frac{h\,c}{k_B\,\lambda_{21 cm}}= 0.068\, {\rm K}$.
Assuming the collision of hydrogen with itself, electron and proton, the total collisional coupling can be written as~\cite{Pritchard-Loeb}
$$x_c = x_c^{HH}\,+\,x_c^{He}\,+\,x_c^{Hp}$$
\begin{equation}
    = \frac{T_\star}{A_{10}T_\gamma}\left[k_{10}^{HH}(T_k)\,n_H\,+k_{10}^{He}(T_k)\,n_e\,+k_{10}^{Hp}(T_k)\,n_p\,\right]
\end{equation}
where $n_H,\,n_e,\,n_p$ are the number densities of neutral hydrogen, electron and proton respectively, and $k_{10}^{HH},\,k_{10}^{He},\,k_{10}^{Hp}$ are the scattering rate between hydrogen atoms with itself, electron and proton respectively. The values of scattering rate for $H-H$, $e-H$, and $p-H$ with respect to gas temperature $T_k$ are listed in tabular form in Ref.~\cite{Cosmo-low-freq-data1,data2}.

\subsection{Wouthuysen Effect}
In this subsection, we review the Wouthuysen-Field effect and the mathematical form of the coefficients describing it. The Wouthuysen--Field effect, named after Siegfried Adolf Wouthuysen and George B. Field, becomes particularly important during the cosmic dawn, when the first luminous sources begin to form. The emergence of early stars and quasars during the epoch of reionization leads to the production of abundant Ly$\alpha$ photons in the universe. These Lyman-$\alpha$ photons might have interacted with the neutral hydrogen, resulting in the spin flip of the electron in the hydrogen atom and therefore in the emission of 21-cm photons. According to the selection rule, the hydrogen atom in the hyperfine state can make a transition as $\Delta F = 0,\pm1$ with $0 \nrightarrow 0$. The process takes place when the hydrogen atom in the s-orbital level of the hyperfine triplet state absorbs the Lyman-$\alpha$ photon and goes to the p-orbital level of the triplet state. After some time following the emission of Lyman-$\alpha$ photon, the hydrogen may come down to the s-orbital level of the singlet state instead of the triplet state, resulting in the spin flip of the electron and emission of the 21-cm signal.  Similarly, the 21-cm photons can also be emitted from the absorption of a Ly$\alpha$ photon by the neutral hydrogen in the hyperfine singlet state $1S$ returning to the $1S$ triplet state. 
\begin{figure}
    \centering
    \includegraphics[width=\linewidth]{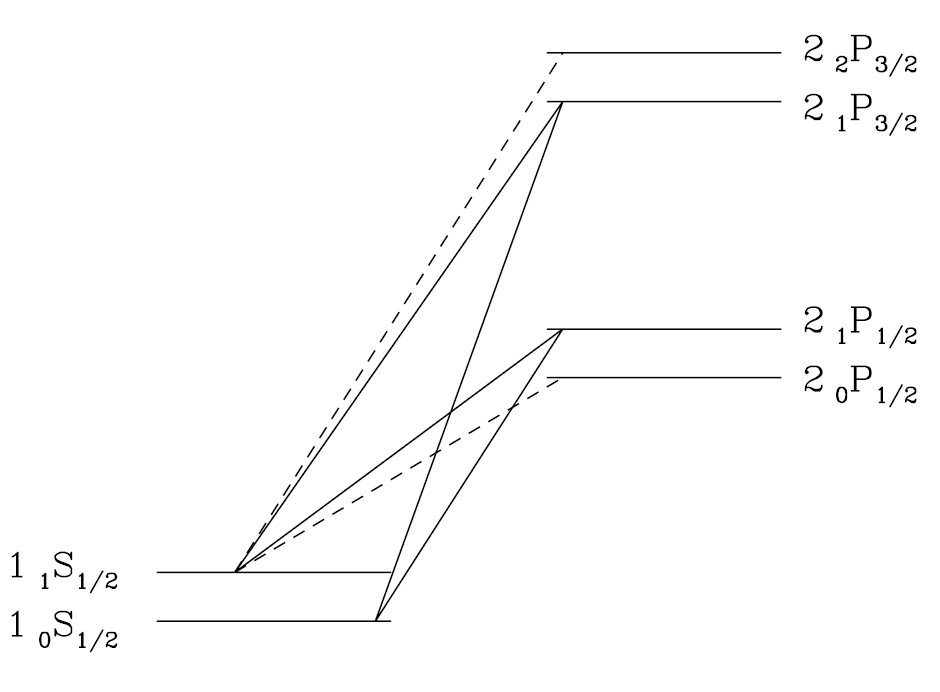}
    \caption{Transitions from the hyperfine splitting resulting in the absorption or emission of 21-cm photons as described by the Wouthuysen Field Effect. From~\cite{Pritchard-Loeb}.}
    \label{fig:placeholder}
\end{figure}

The coupling coefficient for the Wouthuysen Effect can be written in the form of a temperature difference between two hyperfine states ($F=0$ and $F=1$) as~\cite{Pritchard-Loeb}
\begin{equation}
    x_\alpha = \frac{4\,P_\alpha}{27\,A_{10}}\,\frac{T_\star}{T_\gamma},
\end{equation}
where $P_\alpha$ is the scattering rate of Lyman-$\alpha$ photons. We use the $tanh$ parameterization approach to calculate the Ly$\alpha$ coupling coefficient adopted from Ref.~\cite{Ly-alpha-x1,Ly-alpha-x2,Ly-alpha-x3}. The Lyman-$\alpha$ coupling coefficient can be written as
\begin{equation}
    x_\alpha \equiv \frac{2\,A_\alpha(z)}{1+z},
\end{equation}
where the Ly$\alpha$ coupling amplitude $A_\alpha(z)$, according to the $tanh$ model can be quantified as
\begin{equation}
    A_\alpha(z) = A_\alpha (1\,+\,tanh\,[(z_{\alpha\,0} - z)/\Delta z_\alpha]),
\end{equation}
where $z_{\alpha 0}$ is the pivot redshift, $\Delta z$ is the duration and $A_\alpha$ is the step height are free parameters changing the Ly$\alpha$ coupling coefficient $x_\alpha$. We use the same fiducial values of the parameters as given in Ref.~\cite{Ly-alpha-x1} given as 
$$\{A_\alpha,\, z_{\alpha0}, \,\Delta z_\alpha\} = \{100,\, 17,\,2\}$$.

\section{Evolution of the Kinetic Gas Temperature}
\label{gas-temp-standard}

In this section, we discuss the evolution of the kinetic gas temperature in standard cosmology, taking into account the standard spin-flip processes described in the aforementioned sections, as well as possible modifications arising from exotic physics beyond the Standard Model. The kinetic temperature $T_k$ directly affects the spin temperature and consequently modifies the 21-cm brightness temperature, as can be seen from eq.~\eqref{spin-temp}.
\subsection{Gas Temperature Evolution in standard cosmology}
In standard cosmology, the evolution of the kinetic gas temperature is governed primarily by the adiabatic cooling due to the expansion of the Universe and by the thermal coupling between the gas and CMB photons through Compton scattering. Accordingly, the redshift evolution of the gas temperature can be written as
\begin{equation}
    \frac{dT_k}{dz} = \frac{2T_k}{1+z}+\frac{\Gamma_C}{(1+z)H}(T_k-T_\gamma),
    \label{Tk1}
\end{equation}
where $H$ is the Hubble parameter at the redshift of interest, $z$ is the redshift and $\Gamma_C$ is the Compton scattering rate given as
\begin{equation}
    \Gamma_C = \frac{8\,\sigma_T\, a_r\,T_\gamma^4}{3\,m_e\,c}\frac{x_e}{(1+f_{He}+x_e)},
\end{equation}
where $\sigma_T$ is the Thompson scattering cross section, $a_r$ is the Stefan-Boltzmann constant, $m_e$ is the mass of an electron, $ c$ is the speed of light, and $f_{He} = n_{He}/n_H=0.08$ is the fraction of helium atoms with $n_{He}$ and $n_H$ being the number density of helium and hydrogen. Furthermore, the kinetic gas temperature can also be affected by heating of the intergalactic medium by the CMB, which can be quantified as~\cite{Tk-formula2}
\begin{equation}
    \frac{dT_k}{dz} = \left.\frac{dT_k}{dz}\right|_{eq.\ref{Tk1}} - \frac{x_{HI}\, A_{21}}{2\,H\,(1+z)}x_{CMB}\left(\frac{T_\gamma}{T_s}-1\right)T_\star
    \label{Tk2}
\end{equation} 
where $x_{CMB}=\frac{1}{\tau_{21}}(1-e^{-\tau_{21}})$ is the 21-cm optical depth which can be expressed as~\cite{Tk-formula2}
\begin{equation}
    \tau_{21} = \frac{3}{32\pi}\frac{n_H\,x_{HI}\,A_{10}\,\lambda_{21}^3}{H}\frac{T_\star}{T_s}
\end{equation}
\begin{equation}
    \approx 8.1\times 10^{-2}\,\, x_{HI}\left(\frac{1+z}{20}\right)^{3/2}\left(\frac{10\,K}{T_s}\right)
\end{equation}
This is the standard expression for $T_k$ to calculate the global brightness temperature when considering only the general effects. 
\subsection{Gas Temperature Evolution with Exotic Energy Injection}
\label{gas-temp-exotic}
In this subsection, we study the inclusion of additional energy injection terms arising from exotic processes beyond standard cosmology. The energy deposited into the intergalactic medium (IGM) from such processes can primarily proceed through three channels: heating (`heat`), ionization (`ion`), and excitation (`exc`). The corresponding energy deposition rate~\cite{energy-dep1} into a given channel  can be parametrized as:
\begin{equation}
    \left(\frac{dE}{dVdt}\right)_{dep} = f_c\left(\frac{dE}{dVdt}\right)_{inj},
\end{equation}
where $f_c$  for each channel is the deposition fraction for each channel, which can be quantified in terms of the fraction of ionized hydrogen as
\begin{equation}
    f_{ion}\sim f_{exc}\sim \frac{1-x_e}{3},\,\,\,f_{heat} = \frac{1+2x_e}{3},
\end{equation}
Therefore, the final expression of kinetic heating, including the energy into the gas from any exotic phenomenon, can be expressed as
\begin{equation}
    \frac{dT_k}{dz} = \left.\frac{dT_k}{dz}\right|_{eq.\ref{Tk2}} + \frac{dT_k^{inj}}{dz},
    \label{Tk3}
\end{equation}
where the second term corresponds to the kinetic heating of the IGM gas from exotic processes like DM annihilation, neutrino interactions, PBH effects, etc. Here, we will study the effect of extra photons coming from interactions of neutrinos originating from PBHs with background neutrinos. The energy injection from exotic processes can be formulated as~\cite{energy-dep1,energy-dep2}
\begin{eqnarray}
   &&  \frac{dT_k^{inj}}{dz} \equiv \frac{1}{H\,(1+z)}\frac{2\,\dot{Q}}{3k_B(1+f_{He}+x_e)n_H},\nonumber\\
  &&\dot{Q}\equiv f_{heat}(z)\left(\frac{dE}{dVdt}\right)_{inj},
    \label{energy-dep-og}
\end{eqnarray}
where $\dot{Q}$ is the exotic heating rate defined as the rate of change of energy per volume per time and $f_{heat}(z)$ is the deposition fraction from the heating channel. Similarly, we can also find the evolution of the fraction of ionized hydrogen $x_e$ with respect to redshift $z$ as~\cite{xe-formula-1}
\begin{equation}
    \frac{dx_e}{dz} = \frac{1}{H(1+z)}\left[R_s(z)-I_s(z)\right] + \left.\frac{dx_e}{dz}\right|_{\rm exotic},
    \label{dxe-1}
\end{equation}
where $\left.\frac{dx_e}{dz}\right|_{\rm exotic}$ is the evolution of ionization fraction from exotic processes, and $R_s$ and $I_s$ are the recombination and ionization rate of transition of ionized to neutral hydrogen (and vice-versa) given as 
\begin{equation}
    R_s(z) = \mathcal{P}\times[\alpha_H\,x_e\,x_{HII}n_H],
\end{equation}
\begin{equation}
    I_s(z) = \mathcal{P}\times\left[\beta_H\,(1-x_{HII})\,e^{\frac{E_\alpha}{k_B\,T_\gamma}}\right],
\end{equation}
where  $E_\alpha = h\nu_\alpha = 3\mathscr{R}/4$ is the energy corresponding to Ly$\alpha$ transition~\cite{xe-formula-1} with $\mathscr{R}=13.6\,eV$ being the Rydberg constant. The quantity $P$ is known as Peebles' coefficient and represents the probability that a hydrogen atom in the excited state reaches the ground state before being photoionized. It is given by~\cite{xe-formula-1,xe-formula-2,energy-dep1} 
\begin{equation}
    \mathcal{P} = \frac{1+K_H\Lambda_H\,n_H(1-x_{HII})}{1+K_H(\Lambda_H+\beta_H)\,n_H(1-x_{HII})},
\end{equation}
where $K_H = \pi^2/(E_\alpha^3 H)$ is the redshift parameter associated with Ly$\alpha$ photons due to the expansion of the Universe, and $\Lambda_H = 8.22~\mathrm{s}^{-1}$ denotes the decay rate for the $2S \rightarrow 1S$ transition in the hydrogen atom. Assuming the free electron fraction to be equal to the ionized hydrogen fraction, i.e., $x_e \equiv x_{\mathrm{HI}}$, the case-B recombination coefficient $\alpha_H$ and the photoionization rate $\beta_H$ can be expressed as
\begin{eqnarray}
 &&   \alpha_H(T_k) = F\times 10^{-19} \left(\frac{a\,t^b}{1+c\,t^d}\right) \,m^3s^{-1} \nonumber\\
&& \hskip -0.5in \beta_H(T_k) = \alpha_H(T_k)\left(\frac{2\pi\,m_e\,k_B\,T_k}{h_p^2}\right)^{3/2}\,e^{E_\alpha/k_BT_k},
\end{eqnarray}
where values of the parameters are $F=1.14,\,a=4.309\,,b=-0.6166\,,c=0.6703\,,d=0.53\,,t = \frac{T_k}{10^4\,K}$. The fractional ionization evolution due to the exotic process (in our case PBH) present in the eq.~\ref{dxe-1} can also be written in terms of energy deposition terms, just like $T_k$ as,
\begin{equation}
\begin{aligned}
    \left.\frac{dx_e}{dz}\right|_{\rm exotic} &= -\frac{1}{H(1+z)}\frac{1}{n_H}\times\\
    &\quad \left[\frac{1}{E_0}\left.\frac{dE^{dep}}{dVdt}\right|_{ion} \,+ \, \left.\frac{1-\mathcal{P}}{E_\alpha}\frac{dE^{dep}}{dVdt}\right|_{exc}\,\right]
\end{aligned}
\label{modifiedxe}
\end{equation}
We will use the evolution equations for $T_k$ and $x_e$, including the additional energy injection terms, to study the effects of neutrino self-interactions and determine the resulting global 21-cm brightness temperature $T_b$.

\section{PBH emission into neutrinos}
\label{PBH-neutrinos}


Primordial black holes (PBHs), formed through the gravitational collapse of overdense regions in the early Universe, can evaporate into various particle species such as photons, neutrinos, dark matter particles, and other Standard Model degrees of freedom through Hawking radiation. In this section, we evaluate the neutrino flux produced from PBH evaporation and study its impact on the 21-cm brightness temperature through additional energy injection into the intergalactic medium. The mass of a PBH is related to its formation time and can be expressed through the relation~\cite{OUR-PAPER,carr2021constraints}
\begin{equation}
    m_{\rm {BH,0}} \sim 10^{15}\,\left(\frac{t_F}{10^{-23}\,s}\right)\,g
\end{equation}
where ${\mathrm m_{BH},0}$ denotes the initial mass of the PBH at the time of its formation, $t_F$. From this relation, one finds that a PBH with mass $\sim 10^{15}\,\mathrm{g}$ would have formed approximately $10^{-23}\,\mathrm{s}$ after the Big Bang, whereas a PBH with mass $\sim 1\,M_\odot$ would have formed around $10^{-5}\,\mathrm{s}$ after the Big Bang. As the PBH evaporates through Hawking radiation, its temperature increases with time and is inversely proportional to the PBH mass $M_{\mathrm{BH}}$~\cite{OUR-PAPER,HEandUHEchina,NonCDMpaper}.
\begin{equation}
    T_{\rm BH} = \frac{M_p^2}{8\pi M_{\rm {BH}}},
\end{equation}
where ${\rm M_p} = 1.22 \,\times \,10^{19}$ GeV is the Planck mass.
From this relation, it follows that as the PBH evaporates, its mass continuously decreases, leading to a substantial increase in its temperature. Consequently, during the final stages of evaporation, the PBH emits highly energetic particles, with energies potentially approaching the Planck scale. The Hawking radiation emission rate for a particle species $i$ from a Schwarzschild PBH is given by
\begin{equation}
    \frac{d^2N_i}{dt\,dE} = \frac{g_i}{2\pi}\frac{\gamma_{grey}}{exp(E/T_{BH})\pm 1}
    \label{Hawking-rate}
\end{equation}

where $g_i$ is the multiplicity factor associated with the particle species $i$, and $\gamma_{\mathrm{grey}} \simeq 0.2$ represents the greybody factor, which depends on both the energy of the emitted particle and the PBH mass. Although the energy of the emitted particles increases as the PBH evaporates, the total number of emitted particles, and hence the particle flux, decreases with time due to the reduction in the PBH mass. To compute the differential number spectrum of particles of species $i$ emitted from the PBH with respect to the particle energy, we integrate Eq.~\eqref{Hawking-rate} over the lifetime of the PBH, from its formation time to the time of complete evaporation~\cite{HEandUHEchina}:
\begin{equation}
    \frac{dN_i}{dE} = \int^{t_{ev}}_{t_F} \frac{d^2N_i}{dt\,dE} dt
    \label{Hawking-rate-int}
\end{equation}
Assuming the emitted particle species $i$ to be massless, which in our case corresponds to neutrinos, and employing the Boltzmann approximation, Eq.~\eqref{Hawking-rate-int} can be solved analytically to obtain the neutrino spectrum over a given energy interval~\cite{HEandUHEchina}:
\begin{equation}
    \frac{dN_i}{dE} \approx \frac{243\,g_i\,\tau_{BH}}{16\,\pi^3} \, \left[\frac{1}{x^3}-\frac{\Gamma(5,x)}{24\,x^3}\right],
\end{equation}
where $\Gamma(5,x) = e^{-x}\,(x^4+4x^3+12x^2+24x+24)$ is the incomplete gamma function, and $\tau_{BH}$ denotes the lifetime of the PBH. We have also defined the dimensionless variable~$\mathrm {x \equiv {E}/{T_{BH,0}}}$\footnote{Note that, considering the expansion of the universe, $x = \frac{E}{T_{\rm BH,0}}\frac{a_t}{a_{ev}}$, where $a_t$ is the scale factor at any given redshift and $a_{ev} = 2.5\times10^{-31}\left(\frac{m_{\rm BH, 0}}{M_p}\right)^{3/2}$ is the scale factor at which the PBH is completely evaporated.}. 
Similarly, we can also evaluate the rate of change of PBH mass with evaporation by integrating eq.~\ref{Hawking-rate} multiplied with energy E over the interval E~\cite{NonCDMpaper,OUR-PAPER},
\begin{equation}
    \frac{dM_{\rm BH}}{dt} = -\sum_j \int_0^{\infty}E \frac{d^2N_i}{dt\,dE} dE = -e_T\frac{M_p^4}{m^2_{\rm BH,0}},
    \label{Hawking-rate-int2}
\end{equation}
where $g_{*BH} = 108.5$ and 
\begin{equation}
    e_T = \frac{27}{4}\frac{g_{*\rm BH}}{30720\pi}.
\end{equation} 
Substituting into eq.~\ref{Hawking-rate-int2}, we get the PBH mass evolution with time,
\begin{equation}
    \label{mass evolution}
    M_{\rm BH}(t) = m_{\rm BH,0}\left(1-\frac{(t-t_F)}{\tau_{\rm BH}}\right)^{1/3},    
\end{equation}
with $\tau_{\rm BH}$ being the lifetime of PBH which can be calculated as
\begin{equation}
    \label{PBH_lifetime}
    \tau = \frac{1}{3e_T}\frac{m_{\rm BH, 0}^3}{M_p^4}.
\end{equation}
From eq.~\ref{PBH_lifetime}, can observe that any PBH of mass $m_{\rm BH,0}<10^{15}$g would have been completely evaporated before the present universe. Since we are concerned about the 21-cm brightness temperature in the intermediate universe at lower redshift $z\sim 20$, we will only be considering the heavy mass PBH of mass $m_{\rm BH,0}>10^{14}$g. Any PBH of mass lower than this would have been evaporated very early, emitting particles including neutrinos, but as they would have been emitted early, the redshifted energy of neutrinos would be less enough that they might not have any effect in the deviation of brightness temperature beyond the standard processes.  
\subsection{Differential flux of neutrinos emitted from PBH}
To calculate the flux of neutrinos emitted from PBHs, one must solve the Boltzmann equation, which incorporates the effects of Hubble expansion, cosmological redshift, interactions with the thermal plasma, and washout effects in the high-energy tail of the distribution. The Boltzmann equation for the phase-space distribution function $f_i(t,p)$ of particle species $i$ can be written as~\cite{HEandUHEchina}
\begin{equation}
    \rm \left[\frac{\partial}{\partial t}-H\,p\frac{\partial}{\partial p}\right]f_i(t,p) \approx \Gamma_{i,\rm prod} - \Gamma_{i,abs}f_i
\end{equation}
where $f_i(t,p)$ is the phase-space distribution function of particle \textit{i}, in our case $\nu$, $\Gamma_{i,\rm prod}$ and $\Gamma_{i,\rm abs}$ are the parameters taking into account the production of particle $i$ and the absorption rate in the plasma, $H = \dot{a}/a$ is the Hubble parameter, with $a$ being the scale factor. For the case of neutrinos, we take the contribution of $\Gamma_{i,\rm abs}$ to be negligible. Then we, can compute the phase-space distribution as~\cite{HEandUHEchina}
\begin{equation}
    f_i(t,p) = \int^a_0 \frac{\Gamma_{i,\rm prod}(a',p')}{H(a')\,a'}da'
\end{equation}
where $p' \equiv pa/a'$. The $\Gamma_{i,\rm prod}$ can we expressed as
\begin{equation}
    \Gamma_{i,\rm prod}^{(\rm PBH\rightarrow \,i)} \approx n_{\rm BH}\frac{(2\pi)^3}{4\pi p^2}\frac{d^2N_i}{dtdp}
\end{equation}
where $n_{\rm BH}$ is the number density of the PBHs, which for the expanding universe can be written as
\begin{equation}
    n_{\rm BH} = n_{\rm BH,0}\frac{a_F^3}{a^3},\,\,\,\, a\in[a_F,a_{ev}],
\end{equation}
with
\begin{equation}
    n_{\rm BH,0}  =\frac{3\,\beta\,\gamma^2\,M_{p}^6}{32\,\pi\,m_{\rm BH,0}^3}.
\end{equation}
Now, we define a new parameter $\beta$ which denotes the initial fraction of PBH expressed as~\cite{HEandUHEchina,Carr-fPBH-constraint}
\begin{equation}
    \beta \equiv \left.\frac{\rho_{\rm BH,0}}{\rho_{\rm tot}}\right|_{t=t_F},
\end{equation}
where $\rho_{\mathrm{BH},0}$ and $\rho_{\mathrm{tot}}$ denote the energy densities of PBHs and the total energy density of the Universe at the PBH formation time $t_F$, respectively. Using this relation, one can determine the phase-space distribution function for massless neutrinos emitted from PBH evaporation as~\cite{HEandUHEchina}
\begin{equation}
    f_\nu|_{t\rightarrow t_{ev}} \approx f_{\nu,0}\,\left[\frac{1-\Gamma(5,x)/24}{x^5}+f_{\nu,\rm IR}\right],
    \label{phase-space}
\end{equation}
with $f_{\nu,0}$ being the initial phase space distribution function
\begin{equation}
    f_{\nu,0} \equiv \frac{3^6}{4}\,\sqrt{3\,\gamma\, g_{BH}}\beta\,g_\nu\frac{M_p}{m_{\rm BH,0}},
    \label{fi0}
\end{equation}
and $f_{\nu,\rm IR}$ being the IR contribution of $\nu$ which is dominant at low energy tail ($x<<1$)
\begin{equation}
    f_{\nu,\rm IR} = \frac{1}{36x}[log(e^x+1)-x-k_{+}xe^{-x}], \,\,\, ({\rm Fermi-Dirac})
    \label{fiIR}
\end{equation}
where $g_{\rm BH}=\frac{m_{\rm BH,0}^3}{3M_p^4\tau_{\rm BH}}\sim 7.5\times 10^{-3}$ and $k_+ = \frac{3}{4}-\frac{3\zeta(3)}{16}$. Substituting eq.~\ref{fi0} and \ref{fiIR} into eq.~\ref{phase-space}, we will get the phase-space distribution function of neutrinos. From the phase-space distribution, we can evaluate the differential neutrino flux as~\cite{HEandUHEchina,OUR-PAPER}
\begin{equation}
    \frac{d\Phi_\nu}{dE_\nu} = \frac{E_\nu^2}{2\pi^2}f_\nu,
    \label{dif-neutrino-flux}
\end{equation}
where $\Phi_\nu$ is the number of neutrinos etected per unit area per unit time and $E_\nu$ is the energy of the neutrinos. 
Therefore, varying the initial PBH mass $m_{\mathrm{BH},0}$ and the initial PBH abundance parameter $\beta$ modifies the neutrino flux, which subsequently affects the corresponding energy injection rate into the intergalactic medium.
\section{Energy deposition from radiative neutrino scattering}
\label{PBH-energy-deposition}
In this section, we compute the additional energy deposition rate arising from photons produced through interactions between neutrinos emitted by PBHs and the relic neutrinos of the cosmic neutrino background (C$\nu$B). The resulting energy injection heats the IGM during the intermediate stages of cosmic evolution and consequently modifies the global 21-cm brightness temperature. Such radiative scattering processes can arise in several extensions of the Standard Model, where a new mediator interacts with both neutrinos and charged leptons. Since this work represents the first study exploring the impact of these interactions on the global 21-cm signal, we restrict our analysis to a simplified toy-model framework. In particular, we consider a low-energy scenario in which a real singlet scalar field couples to neutrinos as well as charged leptons. The relevant interaction terms can be written as
\[
\mathcal{L} \supset g_{\nu_i}\,\phi\,\nu_i\nu_i + g_{l_i}\,\phi\,\bar{l_i}l_i \, ,
\]
where $i=e,\mu,\tau$ labels the three neutrino flavors. Throughout this work, neutrinos are assumed to be Majorana particles. Accordingly, Weyl notation is adopted for neutrino interactions with the scalar mediator, while Dirac notation is used for charged leptons. Similar interactions may also arise in models involving an additional vector mediator such as a new $Z$ boson. In general, the coupling strengths $g_{\nu_i}$ and $g_{l_i}$ depend on the underlying particle physics realization. However, for simplicity, assume $g_{l_i}=g_{\nu_i}\equiv g_i$ in our toy model.
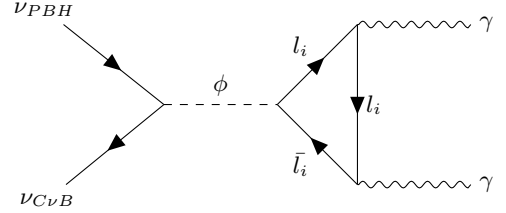
\begin{figure}[H]
    \centering
    \input{figures/feynman-diagram}
    \caption{Feynman diagram for the secondary production of photons from neutrino self interaction between PBH emitted neutrino and neutrinos from C$\nu$B.}
     \label{fig-feynmann}
\end{figure}
The interactions described in the Lagrangian can induce photon production through one-loop processes involving charged leptons and the scalar mediator, as illustrated in Fig.~\ref{fig-feynmann}. The photons generated through these radiative scatterings can deposit energy into the intergalactic medium (IGM), thereby altering its thermal evolution and consequently affecting the global 21-cm signal during the dark ages and cosmic dawn. The cross-section for the one-loop radiative scattering process shown in Fig.~\ref{fig-feynmann} is given by~\cite{Dhuria:2023itq}
\begin{equation}
\label{scatteringcrosssection}
    \sigma = \frac{81\alpha^2 s}{4\pi^3}
    \frac{g_i^4}{(s-m_\phi^2)^2+m_\phi^2\Gamma_\phi^2}
    \times
    \left|1+ Q_i^2m_i^2 C_0^\gamma\right|^2 \, .
\end{equation}
where $C_0^\gamma$ denotes the scalar Passarino--Veltman function,
\begin{equation}
\label{Passarino-Veltman Function}
    C_0^\gamma(s,m_i)=\frac{1}{2s}
    \ln^2\left(
    \frac{\sqrt{1-4m_i^2/s}-1}
    {\sqrt{1-4m_i^2/s}+1}
    \right) \, .
\end{equation}
In the above expressions, $g_i$ represents the self-interaction coupling for a given neutrino flavor, while $m_i$ corresponds to the mass of the associated charged lepton. The Mandelstam variable is defined as $s = 2m_{\nu_i}E_{\nu_h}$, where $m_{\nu_i}$ is the mass of the active neutrino and $E_{\nu_h}$ denotes the energy of the high-energy neutrino. The parameter $m_\phi$ is the mass of the scalar mediator, and $\Gamma_\phi = \frac{g_i^2m_\phi}{4\pi}$
is the decay width of the mediator. The scattering cross-section exhibits resonant enhancement when the condition $E_{\nu_h}\approx \frac{m_\phi^2}{2m_{\nu_i}}$ is satisfied. Using the cross-section given in Eq.~\eqref{scatteringcrosssection}, we now investigate the impact of ultra-high-energy neutrino self-interactions on the evolution of the global 21-cm brightness temperature.

The energy deposition arising from photons produced through radiative scattering can be determined from the corresponding interaction rate. For this purpose, we define an effective interaction rate $R'(E)$ associated with the photon production process as
\begin{equation}
    R'(E_\nu,z) = n_1\,n_2\,<\sigma v> \,= n_{\nu,\rm PBH}\, \cdot n_{c\nu B}<\sigma v>
\end{equation}
where $n_{\nu,\mathrm{PBH}}$ and $n_{\mathrm{C}\nu\mathrm{B}}$ denote the number densities of neutrinos emitted from PBH evaporation and relic neutrinos from the cosmic neutrino background, respectively. Here, $v$ represents the relative velocity between the interacting neutrinos, and $\sigma$ corresponds to the scattering cross-section for the interaction process. As the neutrinos emitted from the PBH evaporation are mostly non-thermal, it would be required to conisder the flux of neutrinos instead of number density. Thus, we rewrite the interaction rate as:
\begin{equation}
    \begin{aligned}
        R'(E,z) &= n_{\nu,\rm PBH}(E)\, \cdot n_{c\nu B}(z)\,<\sigma  v> \\
        &= \Phi_{\nu,\rm PBH}(E)\,\cdot  n_{c\nu B}(z)\, \cdot \sigma
    \end{aligned}
\end{equation}
where $\Phi_{\nu,\rm PBH} = (n_{\nu,\rm PBH}(E)\cdot v) $ is the directional neutrino flux originating from the PBH which we can derive from integrating eq.~\ref{dif-neutrino-flux} over energy interval $E_\nu$. Therefore, the energy injection term can finally be written as  
\begin{equation}
    \left.\frac{dE_\nu}{dVdt}\right|_{inj} = \int_{E_{\rm min}}^{E_{\rm max}}  dE_\nu \frac{d\Phi_{\nu,\rm PBH}}{dE_\nu}\,n_{c\nu B}(z) \,\sigma(E_\nu)\,\cdot E_\nu
\end{equation}
From this, we can calculate the energy deposition rate for kinetic heating using eq.~\ref{energy-dep-og}, 
\begin{equation}
    \dot{Q}(E_\nu,z) = f_{heat}\int_{E_{min}}^{E_{max}}  dE_\nu \frac{d\Phi_{\nu,\rm PBH}}{dE_\nu}\,n_{c\nu B}(z) \,\sigma(E_\nu)\,\cdot E_\nu
\end{equation}
where $\rm n_{c\nu B}(z) = n_{c\nu B,0}(1+z)^3$ with $\rm n_{c\nu B,0} \approx 112\,{\rm cm}^{-3}$. By incorporating this energy deposition rate into the evolution equations governing the 21-cm signal, we study its impact on the evolution of the global 21-cm brightness temperature.

\section{Results and Discussion}
\label{results}
In this section, we summarize the impact of PBHs acting as an additional heat injection source in the evolution of the global 21-cm brightness temperature. The heating mechanism arises from the interaction between neutrinos emitted through PBH evaporation and the background relic neutrinos of the cosmic neutrino background. The radiative scattering of these neutrinos produces secondary photons, which subsequently transfer energy to the IGM, thereby modifying its thermal history. 

\subsection{Redshift Evolution of $T_{21}$, $T_{\rm gas}$, and $x_e$}


\begin{figure}[htbp]
    \subfigure[]{
    \centering
    \includegraphics[width=\linewidth]{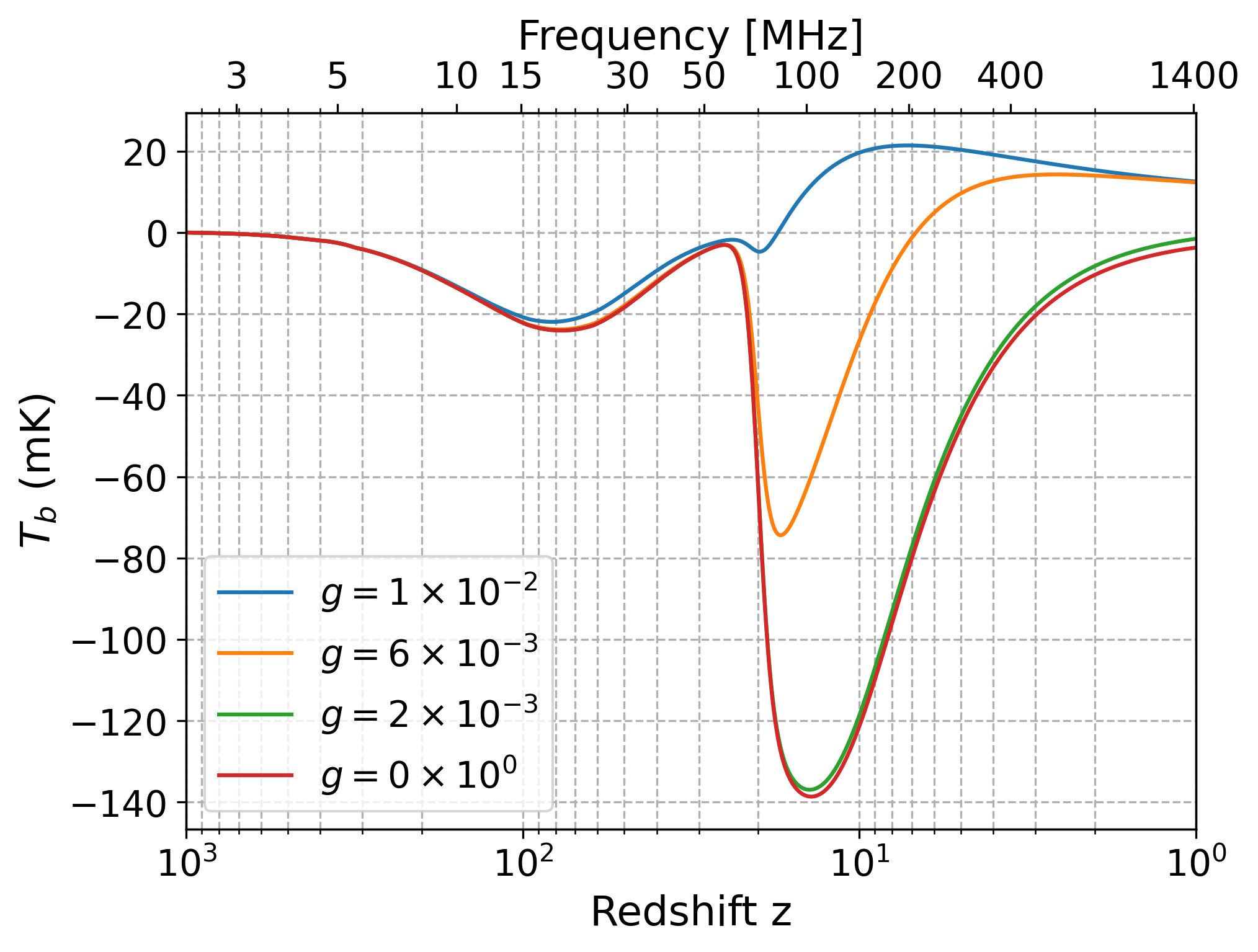}
    \label{Tb-redshift}
    }
    
    \subfigure[]{
    \centering
    \includegraphics[width=\linewidth]{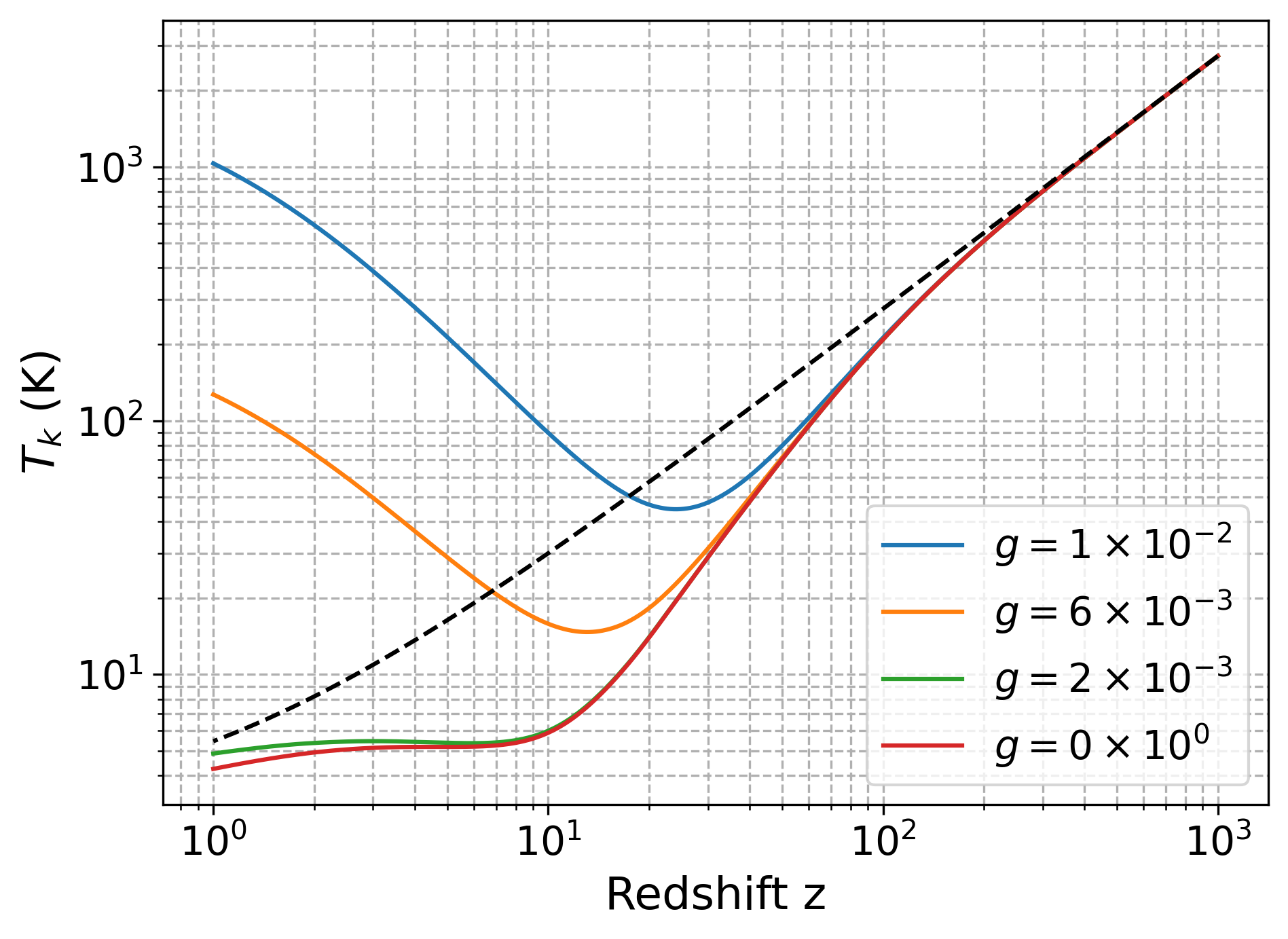}
    \label{Tk-redshift}
    }
   
    \subfigure[]{
    \centering
    \includegraphics[width=\linewidth]{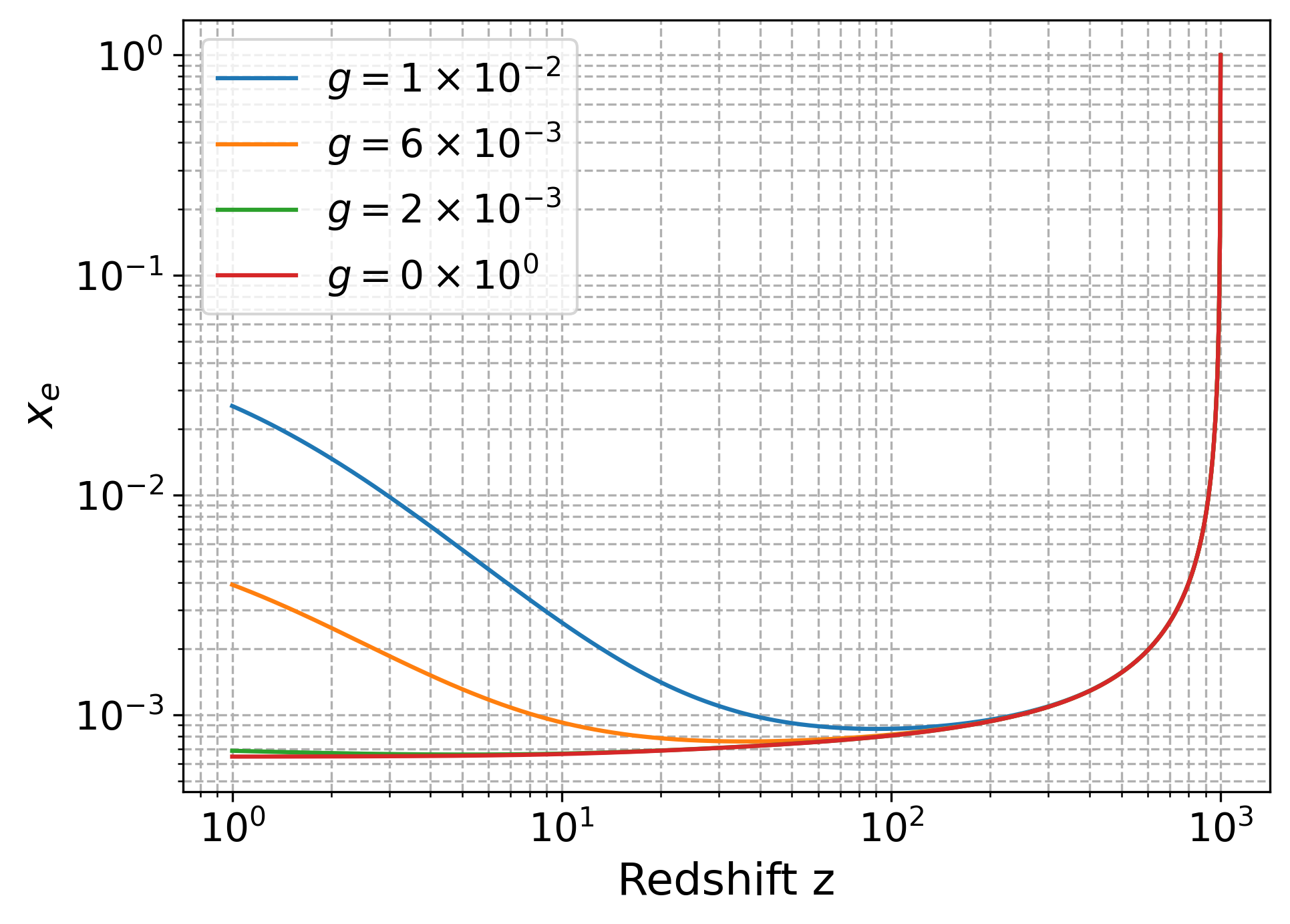}
    \label{xe-redshift}
    }
    \caption{Redshift evolution of the global 21-cm brightness temperature, kinetic gas temperature, and free electron fraction for a fixed mediator mass $m_\phi = 0.1~\mathrm{GeV}$, PBH mass $m_{\mathrm{BH},0}=10^{17}~\mathrm{g}$, and PBH abundance fraction $f_{\mathrm{PBH}}=0.1$, while varying the self-interaction neutrino coupling.}
    \label{Redshift-evolution}
\end{figure}
In this subsection, we first discuss the impact of the secondary photons emitted from neutrino radiative scattering on the profile of 21-cm brightness temperature, gas kinetic temperature, and the free electron fraction.  To study these effects, we analyze the redshift evolution of all three parameters for different values of the neutrino coupling $g$  while fixing fixed viable value of PBH mass $m_{\rm BH,0}$ and PBH fraction $f_{\rm PBH}$. The PBH abundance is parametrized through the fraction $f_{\rm PBH}$ of dark matter residing in PBHs, which can be related to the initial PBH mass fraction $\beta$ following the standard formalism of Ref.~\cite{Carr-fPBH-constraint}. To define the PBH fraction $f_{\rm PBH}$, one has to define a new parameter $\beta'$ which depends on $\beta$ as
\begin{equation}
    \beta' \equiv \gamma^{1/2} \left(\frac{g_{i*}}{106.75}\right)^{-1/4}\, \left(\frac{h}{0.68}\right)^{-2}\,\beta
\end{equation}
Now, the PBH fraction is defined as the ratio of current PBH mass density to that of Cold Dark Matter (CDM) mass density,
\begin{equation}
    f_{\rm PBH} \equiv \frac{\Omega_{\rm PBH}}{\Omega_{\rm CDM}} \approx 4.11\times 10^8 \, \beta'\,\left(\frac{m_{\rm BH,0}}{M_\odot}\right)^{-1/2}
\end{equation}
where $M_\odot$ is the solar mass. We will be using the PBH fraction $f_{\rm PBH}$ to study the constraints and effect on the brightness temperature, gas temperature, and the ionization fraction from the self-neutrino interactions. Other than this, we will be considering the $\tau$-flavor neutrinos in our analysis, as they are comparatively the least constrained by existing astrophysical and cosmological observations. Fig.~\ref{Redshift-evolution} shows the evolution of the differential brightness temperature as a function of redshift $z$. Two characteristic absorption features can be observed around $z\sim100$ and $z\sim10$, corresponding respectively to the post-recombination neutral hydrogen-dominated era and the cosmic dawn/early star formation epoch. Since the cosmic dawn and the epoch of reionization are most sensitive to exotic heating effects, we focus particularly on the constraints at redshift $z\simeq17$. Compared to the standard 21-cm evolution (shown by the red curve corresponding to $g=0$), the inclusion of PBH-induced heating suppresses the absorption signal and drives the brightness temperature toward zero due to the additional heating of the IGM. The magnitude of this suppression depends directly on the neutrino interaction strength. Stronger coupling between the background $\nu_\tau$ and PBH-emitted $\nu_\tau$ enhances the energy transfer to the IGM, resulting in larger gas heating and consequently weaker 21-cm absorption.

Figure~\ref{Redshift-evolution}(b) illustrates the cosmic evolution of the IGM kinetic temperature. At high redshifts, the gas temperature $T_k$ remains tightly coupled to the CMB temperature $T_\gamma$, after which the two temperatures decouple and evolve independently~\cite{Poulin:2016anj}. In the absence of exotic heating ($g=0$), the gas cools adiabatically below the CMB temperature. However, when PBH-induced heating is included ($g\neq0$), the injected energy gradually increases the gas temperature, and for sufficiently large couplings, the gas temperature can exceed the CMB temperature after $z\sim20$, as indicated relative to the black dashed curve representing $T_\gamma$.

The evolution of the free electron fraction is shown in Fig.~\ref{Redshift-evolution}(c). Similar to the behavior of the brightness and gas temperatures, the PBH-induced heating modifies the ionization history at lower redshifts. The additional energy injection increases the residual ionization fraction relative to the standard cosmological evolution, with stronger deviations appearing for larger neutrino couplings.


\subsection{PBH and Neutrino Interaction Constraints}
\begin{figure}[htbp]
    \centering
    \includegraphics[width=\linewidth]{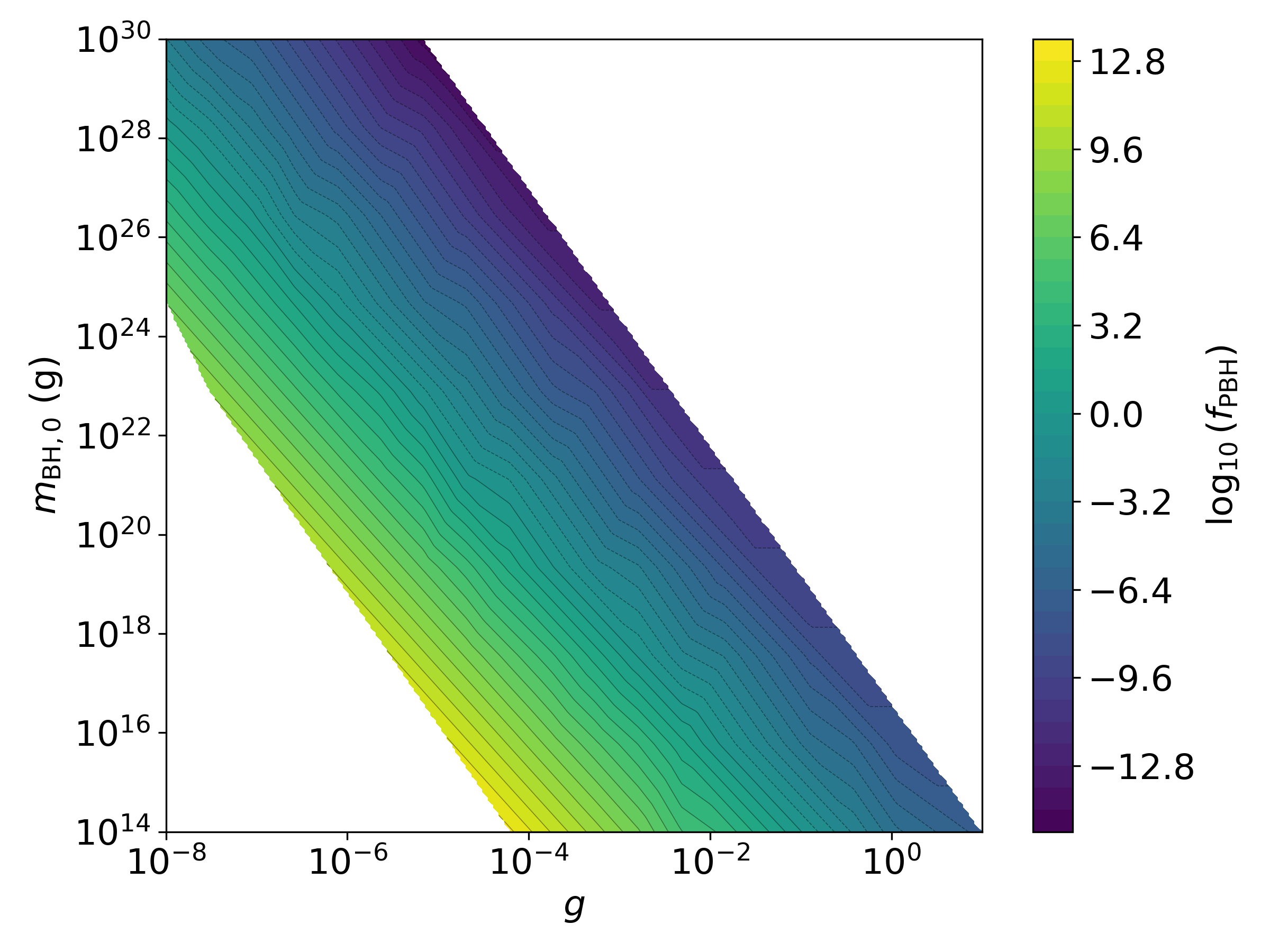}
    \caption{Constraint on PBH fraction for different values of $\tau$-neutrino coupling $g$ for brightness temperature to be $T_b=-100$ mK obtained by fixing $m_\phi=0.1$ GeV at redshift $z=17$. }
    \label{contour-constr}
\end{figure}

\begin{figure}[htbp]
    \centering
    \includegraphics[width=1.05\linewidth]{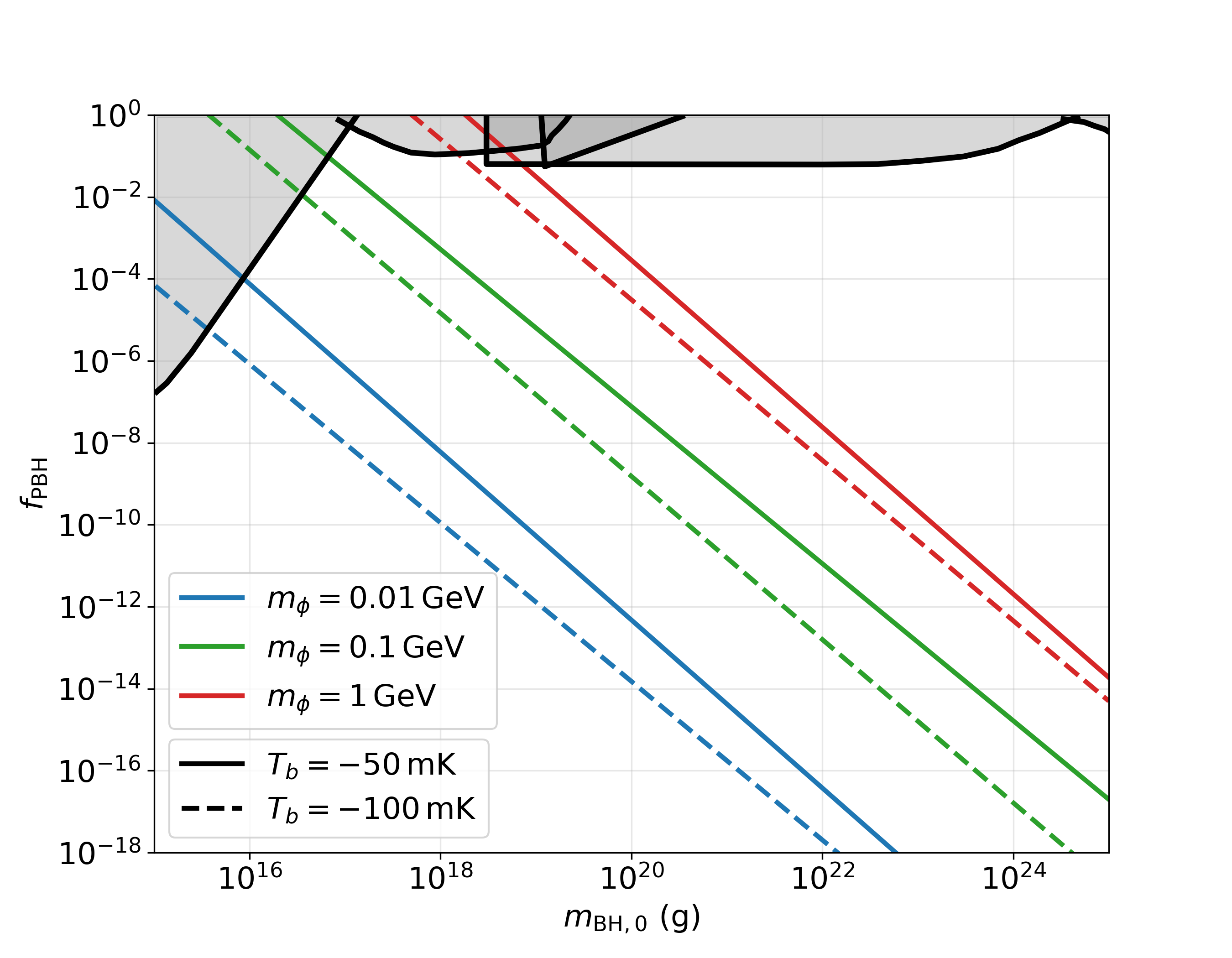}
    \caption{Constraint on the PBH abundance fraction for different values of the $\nu_\tau$ interaction mediator mass, $m_\phi = 0.01,\ 0.1,\ 1~\mathrm{GeV}$, for a fixed $\tau$-neutrino self-interaction coupling $g=0.01$, yielding brightness temperatures $T_b=-100~\mathrm{mK}$ and $T_b=-50~\mathrm{mK}$ at redshift $z=17$. The solid curves correspond to $T_b=-50~\mathrm{mK}$, while the dashed curves correspond to $T_b=-100~\mathrm{mK}$. The shaded region represents the existing observational constraints on the PBH abundance fraction.}
    \label{constraint-mphi}
\end{figure}

 Using the modified global brightness temperature obtained after incorporating neutrino self-interactions, we derive constraints on the PBH abundance fraction for different PBH masses at redshift $z=17$. Figure~\ref{contour-constr} illustrates the parameter space in the plane of PBH mass and neutrino self-interaction coupling for a fixed mediator mass $m_\phi = 0.1~\mathrm{GeV}$. In particular, we determine the PBH abundance fraction required to generate sufficient heating of the IGM such that the global brightness temperature reaches $T_b=-100~\mathrm{mK}$. The analysis covers PBH masses in the range $10^{14}~\mathrm{g} \lesssim m_{\mathrm{BH},0} \lesssim 10^{30}~\mathrm{g}$. From the contour plot, one can infer that larger values of the $\nu_\tau$ self-interaction coupling enhance the energy deposition into the IGM, allowing even relatively low-mass PBHs to produce observable modifications to the 21-cm signal. Figure~\ref{constraint-mphi} shows the corresponding bounds on the PBH abundance fraction obtained by varying the mediator mass $m_\phi$ while keeping the neutrino coupling fixed at $g=0.01$. We find that the constraints become increasingly stringent for heavier PBHs compared to lighter PBHs. Furthermore, increasing the mediator mass suppresses the interaction rate, thereby requiring a larger PBH abundance to inject sufficient energy into the IGM and achieve brightness temperatures of $T_b=-100~\mathrm{mK}$ and $T_b=-50~\mathrm{mK}$.
The results further indicate that stronger PBH-induced heating of the IGM, corresponding to a shallower 21-cm absorption signal with $T_b=-50~\mathrm{mK}$, requires a comparatively larger PBH abundance than the case with $T_b=-100~\mathrm{mK}$. The grey shaded region in the figures represents the parameter space already excluded by existing observational constraints on PBHs~\cite{Carr-fPBH-constraint}. These constraints arise from several independent probes, including extragalactic $\gamma$-ray emission from PBH evaporation~\cite{carr2010new}, femtolensing of $\gamma$-ray bursts~\cite{barnacka2012new}, white dwarf explosions~\cite{graham2015dark}, neutron star capture~\cite{capela2013constraints}, and Kepler microlensing observations~\cite{griest2014experimental}.

\begin{figure}[htbp]
    \centering
    \includegraphics[width=1.05\linewidth]{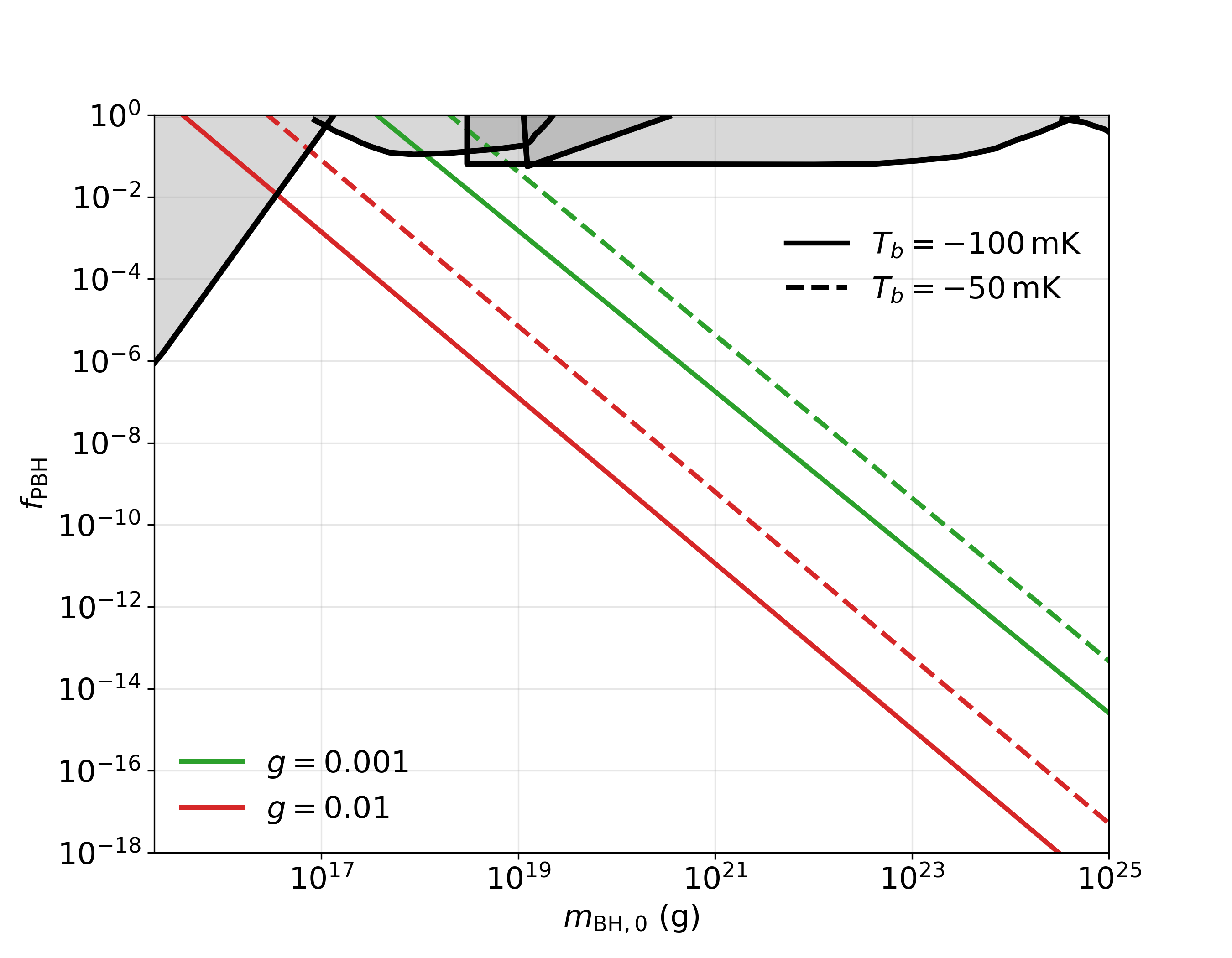}
    \caption{Constraints on PBH fraction at redshift $z=17$ for different interaction coupling $\rm g=0.001,\,0.01$ obtained for different brightness temperature $\rm T_b=-100,-50$ mK  after fixing $m_\phi=0.1$ GeV.}
    \label{constraint-g}
\end{figure}

Similarly, Fig.~\ref{constraint-g} shows the constraints on the PBH abundance fraction obtained by varying the neutrino self-interaction coupling $g$ for a fixed mediator mass $m_\phi=0.1~\mathrm{GeV}$. The contours correspond to parameter regions that produce global brightness temperatures of $T_b=-100~\mathrm{mK}$ and $T_b=-50~\mathrm{mK}$ at redshift $z=17$.

\begin{figure}[htbp]
    \centering
    \includegraphics[width=1.05\linewidth]{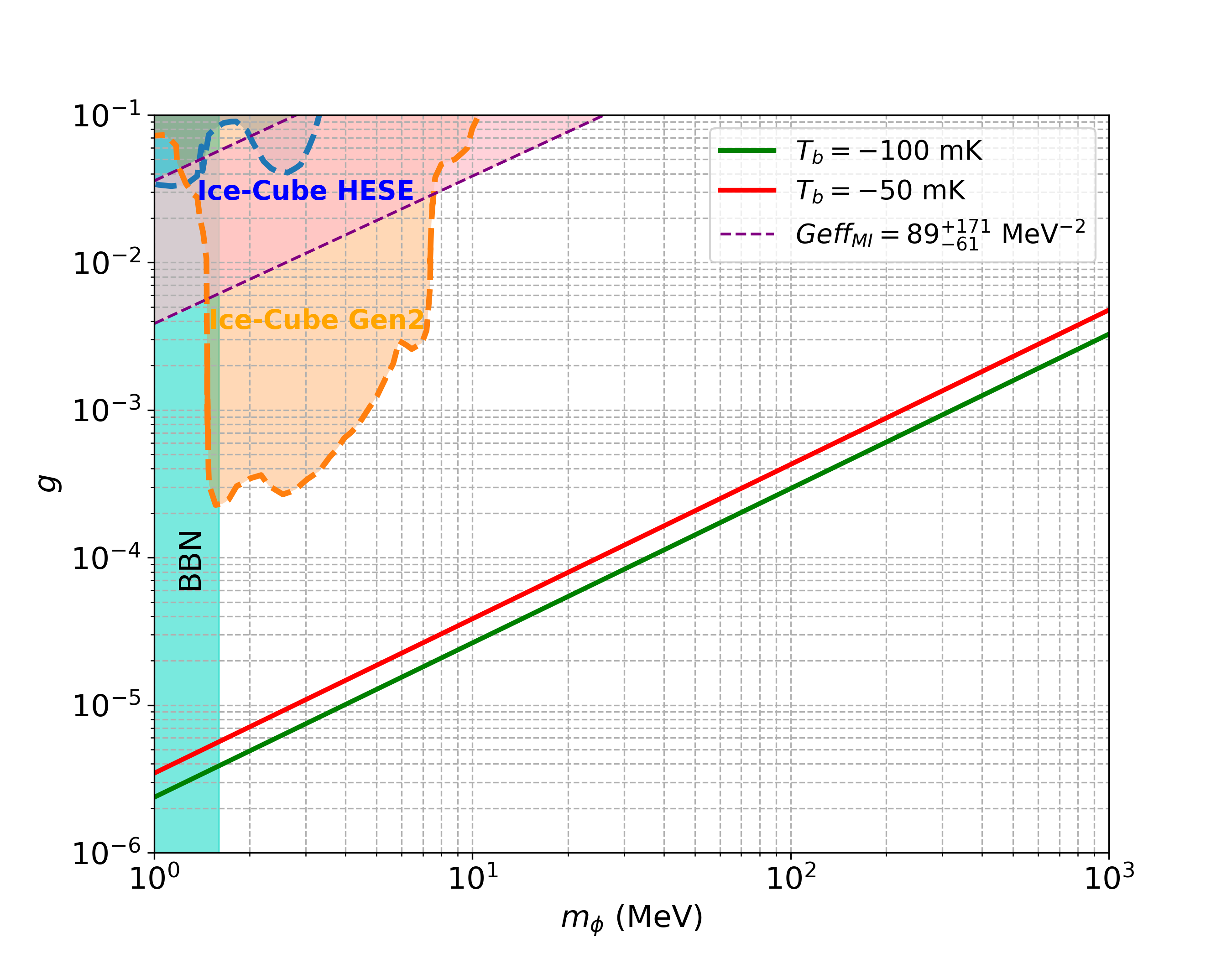}
    \caption{Parameter space for $\nu_\tau$ self-interaction for fixed value of PBH mass $m_{BH,0}=10^{20}$ g and PBH fraction $f_{PBH}=0.01$ such that the total global brightness temperature becomes $T_b=-100,\,-50$ mK. The cyan region represents the parameter space excluded by Big Bang Nucleosynthesis (BBN) constraints~\cite{BBN-constraint}. The blue and orange regions denote the sensitivity reach of current and future IceCube observations, respectively~\cite{Icecube-Gen2-constraint,Icecube-HESE-constraint}, while the purple region indicates the parameter space favored in scenarios addressing the Hubble tension~\cite{Dhuria:2023itq}.}
    \label{g-vs-mphi}
\end{figure}
Finally, Fig.~\ref{g-vs-mphi} illustrates the allowed parameter space in the $(g,m_\phi)$ plane for neutrino self-interactions. In this analysis, we fix the PBH mass and abundance fraction to $m_{\mathrm{BH},0}=10^{20}~\mathrm{g}$ and $f_{\mathrm{PBH}}=0.01$, respectively, consistent with existing observational constraints~\cite{carr2010new}. The allowed regions correspond to combinations of the neutrino self-interaction coupling $g$ and mediator mass $m_\phi$ that yield global brightness temperatures of $T_b=-100~\mathrm{mK}$ and $T_b=-50~\mathrm{mK}$ at redshift $z=17$. These regions therefore represent the viable parameter space consistent with both current PBH bounds and the thermal evolution inferred from the global 21-cm signal. The plot also includes the regions excluded by Big Bang Nucleosynthesis (BBN)~\cite{BBN-constraint}, the shaded parameter space motivated by the absorption features observed in IceCube ultra-high-energy neutrino data~\cite{Icecube-Gen2-constraint,Icecube-HESE-constraint}, and constraints associated with the Hubble tension. We find that the range of neutrino self-interaction couplings probed in our analysis extends to values significantly smaller than those accessible through current IceCube observations and future projections. Consequently, the global 21-cm signal provides a substantially more sensitive probe of such interactions. In particular, even extremely small neutrino self-interaction couplings can produce measurable modifications to the global 21-cm brightness temperature through additional heating of the IGM.

Overall, we conclude that for PBH masses in the range $
10^{15}\,\mathrm{g} \lesssim m_{\mathrm{BH},0} \lesssim 10^{25}\,\mathrm{g}$
and phenomenologically viable values of the PBH abundance fraction $f_{\mathrm{PBH}}$, even very small neutrino self-interaction couplings can induce appreciable heating of the intergalactic gas, thereby affecting the global 21-cm brightness temperature. Our analysis therefore demonstrates that radiative neutrino self-interactions can lead to observable deviations in the 21-cm signal and provide a highly complementary probe of PBH evaporation and beyond Standard Model neutrino physics.

 \section{Summary and Future Plan}
 \label{Summary}
 The global 21-cm signal has emerged as a powerful probe of exotic energy injection processes in the early Universe, providing a unique window into physics beyond the Standard Model. In this work, we explored the effects of neutrinos emitted from primordial black hole (PBH) evaporation on the thermal evolution of the intergalactic medium (IGM) and the resulting modifications to the global 21-cm brightness temperature. In particular, we investigated a scenario in which high-energy neutrinos originating from PBH evaporation undergo radiative scattering with relic neutrinos from the cosmic neutrino background (C$\nu$B), producing photons through loop-induced interactions.

To study this effect, we considered a minimal neutrino self-interaction framework involving a scalar mediator coupled to both neutrinos and charged leptons. Such couplings naturally give rise to one-loop radiative processes in which charged leptons mediate photon production in the final state. The emitted photons subsequently transfer energy to the IGM, altering its thermal history during the dark ages and cosmic dawn epochs. We incorporated the corresponding energy injection terms into the evolution equations governing the gas temperature and ionization fraction to determine the resulting impact on the global 21-cm signal.

The magnitude of the effect depends on several key parameters, including the initial PBH mass $m_{\mathrm{BH},0}$, the initial PBH abundance fraction $\beta$, the neutrino self-interaction coupling $g_i$, and the mediator mass $m_\phi$. Since our primary interest lies in PBHs evaporating at relatively late cosmological times, we restricted our analysis to PBHs with masses $m_{\mathrm{BH},0} \gtrsim 10^{15}\,\mathrm{g}$. We showed that the neutrino flux generated from PBH evaporation can significantly enhance the radiative scattering rate with C$\nu$B neutrinos, thereby increasing the energy deposition into the IGM.

Most previous studies examining the impact of PBH evaporation on the 21-cm signal have focused primarily on direct photon emission from PBHs. In contrast, the mechanism studied here relies on neutrino-induced radiative processes. Since neutrinos are electrically neutral and interact only weakly, they can propagate over cosmological distances without significant attenuation, making them particularly effective probes of high-energy phenomena in the early Universe. Motivated by recent interest in neutrino self-interactions arising from cosmological observations, laboratory bounds, and possible absorption features in IceCube ultra-high-energy neutrino data, we explored the implications of such interactions for the 21-cm signal.

Our analysis demonstrates that, within phenomenologically allowed parameter ranges, the radiative scattering of PBH-induced neutrinos can lead to substantial heating of the IGM gas, thereby increasing the global 21-cm brightness temperature. These results highlight the sensitivity of 21-cm cosmology to neutrino self-interactions and PBH-induced energy injection mechanisms. The framework developed in this work therefore provides a complementary approach to conventional photon-based analyses and opens a new avenue for probing neutrino properties and exotic early-universe physics using future precision measurements of the global 21-cm signal.

 \section{Acknowledgment}
\label{Acknowledgment}
MD would like to acknowledge Havisha Thakur for her contribution during the initial phase of this project as part of her B.Sc. thesis work. MD also acknowledges the organizers of the \emph{Vikram Discussion on Neutrino Astrophysics} (March 2026), held at the Physical Research Laboratory (PRL), Ahmedabad, India, where preliminary results of this work were presented and discussed. MD further acknowledges Prof.~Srubabati Goswami for valuable comments and insightful suggestions.

\bibliographystyle{unsrt}
\bibliography{bibliography}

\end{document}

%% file: figures/feynman-diagram.tex
\begin{tikzpicture}
    \begin{feynman}
        \vertex (a);
        \vertex [above left= of a] (b){$\nu_{PBH}$};
        \vertex [below left= of a] (c){$\nu_{C\nu B}$};
        \vertex [right= of a] (d);
        \vertex [above right= of d] (e);
        \vertex [below right= of d] (f);
        \vertex [right= of e] (g){$\gamma$};
        \vertex [right=of f] (h) {$\gamma$};

        \diagram{
        (b) --[fermion] (a);
        (a) --[fermion] (c);
        (a) --[scalar, edge label={$\phi$}] (d);
        (d) --[fermion, edge label=$l_i$] (e);
        (f) --[fermion,edge label=$\bar{l_i}$] (d);
        (e) --[fermion, edge label=$l_i$] (f);
        (e) --[boson] (g);
        (f) --[boson] (h);
        };
    \end{feynman}
\end{tikzpicture}